 \documentclass[10pt]{aastex63}		%ArXive preprint
 \usepackage{natbib}
\def\ang{\AA}

\def\gapprox{\lower.4ex\hbox{$\;\buildrel >\over{\scriptstyle\sim}\;$}}
\def\lapprox{\lower.4ex\hbox{$\;\buildrel <\over{\scriptstyle\sim}\;$}}

\shortauthors{Aschwanden \& Schrijver}
\shorttitle{Solar-Stellar Connection}

\begin{document}
\renewcommand{\topfraction}{0.95}
\renewcommand{\bottomfraction}{0.95}
\renewcommand{\textfraction}{0.05}
\renewcommand{\floatpagefraction}{0.95}
\renewcommand{\dbltopfraction}{0.95}
\renewcommand{\dblfloatpagefraction}{0.95}

%{\sl  Manuscript, version 2025 Mar 9; draft ... }

\title{ Self-Organized Criticality Across 
	Thirteen Orders of Magnitude in
	the Solar-Stellar Connection }
 
\author{Markus J. Aschwanden \& Carolus, J. Schrijver}
\affil{Lockheed Martin, Solar and Astrophysics Laboratory (LMSAL),
       Advanced Technology Center (ATC),
       A021S, Bldg.252, 3251 Hanover St.,
       Palo Alto, CA 94304, USA;
       e-mail: markus.josef.aschwanden@gmail.com}

\begin{abstract}
The observed size distributions of solar and stellar flares
is found to be consistent with the predictions of the 
fractal-diffusive self-organized criticality (FD-SOC) model, 
which predicts power law slopes with universal constants of 
$\alpha_F=(9/5)=1.80$ for the flux, and
$\alpha_E=(5/3)\approx 1.67$ for the fluence, respectively.  
In this Letter we explore the solar-stellar connection
under this aspect, which extends over an unprecedented dynamic 
range of 13 orders of magnitude between the smallest detected
solar nanoflare event ($E_{min}=10^{24}$ erg) and the largest 
superflare ($E_{max}=10^{37}$ erg) on solar-like G-type stars,
observed with the KEPLER mission. The FD-SOC model predicts
a scaling law of $L \propto E^{(2/9)}$ for the length scale
$L$ as a function of the flare energy $E$, which limits the
largest flare size to $L_{max} \lapprox 0.14 R_\odot$ 
for solar flares, and $L_{stellar} \lapprox 1.04\ R_{\odot}$
for stellar flares on G-type stars. In the overall we 
conclude that the universality of power laws (and their
slopes) is a consequence of SOC properties (fractality, 
classical diffusion, scale-freeness, volume-flux 
proportionality), rather than identical physical
processes at different wavelengths. 
\end{abstract}

\footnote{This paper is dedicated to Carolus J. Schrijver 
	(1958-2024) in memoriam.}  

\section{	INTRODUCTION 				}  

The concept of self-organized criticality (SOC) was originally
proposed by Bak, Tang, \& Wiesenfeld (1987) and has been
used in over 6000 publications since. An interdiscipliary
approach of SOC phenomena is sketched in the book by Bak (1996).
The SOC paradigm can be understood most easily with the sandpile 
analogy.
If you trickle sand grains on top of a sandpile, the
slope (or angle of repose) on the inclines of the sandpile
start to steepen until the system reaches a critical angle
(which is about $38^\circ$ in real-world sandpiles). Every
additional sand grain dropped onto the critically inclined
slope will trigger a small or a large avalanche, which 
is a highly nonlinear process. Bak et al.~(1987) developed
the ingenious method of using a n-dimensional lattice grid
to simulate next-neighbor interactions with a computer code.
The ``critical'' sandpile is driven by the balance between 
the input (infalling sand grains) and the output
(or avalanches running off the boundaries at the bottom
of the sandpile).
If the avalanche duration is shorter than the waiting time
interval between subsequent avalanches, the system is said
to be slowly driven. The sandpile experiments of 
Bak et al.~(1987) produced power law distribution functions
with slopes of $\alpha_S=1.37$ for the avalanche size $S$, and
$\alpha_T=0.92$ for avalanche durations $T$, respectively. 
Extensive literature on this type of numerical SOC 
simulations can be found in Pruessner (2012).

About two decades later, after the execution of numerous SOC 
simulations in discretized n-dimensional space, a paradigm
shift occurred that has been named the {\sl fractal-diffusive
self-organized criticality (FD-SOC)} model (see textbooks by
Aschwanden 2011, 2025). The new approach is motivated by the
dilemma that microscopic real-world simulations are 
computationally expensive, if not impossible, such as in the 
case of bridging microscopic (atomic) scales to macroscopic 
scales (measurable in a ``real-world'' scenario). Moreover,
next-neighbor interactions have to be simulated with iterative
steps, which makes it to a mathematically infinite equation
system. Instead, next-neighbor interactions can be approximated
by the classical diffusion process, and the spatial inhomogeneity
of SOC avalanches can be approximated by fractal dimensions.
This way the power law size distributions can be quantified
with simple (macroscopic) physical scaling laws, which obey
the scale-freeness in their statistical probability distribution
functions. It appears that the FD-SOC model is a suitable
description for many astrophysical (solar and stellar) phenomena,
but has also wide-spead applications in geophysics, biophysics,
medical sciences, as well as in financial physics.
More detailed descriptions of the FD-SOC model can be
found in Aschwanden (2014, 2015), Aschwanden 
et al.~(2016, 2021, 2022), Aschwanden \& Gogus (2025),
while alternative SOC reviews are
given elsewhere (Aschwanden et al.~2013, 2016; Watkins et al.~2016; 
Sharma et al.~2016; McAteer et al.~2016).

The contents of this Letter includes observations (Chapter 2),
data analysis (Chapter 3), discussion (Chapter 4), and
conclusions (Chapter 5).

\section{ 	OBSERVATIONS } 

We use multiple data sets to synthesize a joint size distribution,
selected by their published power law slopes $\alpha_E$ and energy 
ranges $E_{min}, E_{max}$ of their energy (in physical units [erg]), 
not to be confused by their power law slopes of peak fluxes 
$\alpha_F$, which are quantified in physical units of [erg/s].
A synthsized size distribution, as composed by 5 individual
size distributions, is fitted to a theoretically predicted
power law slope, as shown in Fig.~(1). While the power law
slopes are invariant accross 13 orders of magnitude, the
energy definitions of these individual data sets are not 
homogeneous here. 
In the following, we subdivide the data sets into three groups
according to the used energy definitions and wavelength dependence 
(Table 1): (i) thermal
energies, radiating in soft X-rays (SXR), extreme-ultraviolet (EUV),
and ultraviolet (UV); (ii) non-thermal energies (radiating in hard X-rays
and gamma rays), and (iii) bolometric energies, radiating in white light. 

\subsection{	Soft X-Ray Thermal Energies 	}

The smallest solar flare events, also called ``nanoflares'' or
``transient brightenings'',
were sampled by observations with SXT/Yohkoh (Shimizu 1995),
EIT/SOHO (Krucker \& Benz 1998; Benz \& Krucker 2002;
TRACE (Aschwanden et al.~2000; Parnell \& Jupp 2000),
and GOES (Li et al.~2012). 

The physical parameters of nanoflares include
electron temperatures of $T_e \lapprox 2$ MK,
electron densities of $n_e \approx \lapprox 10^9$ cm$^{-3}$, 
(loop) length scales of $L = 2 - 20$ Mm, and thermal energies 
of $E_{th}=10^{24}-10^{26}$ erg (Aschwanden et al.~2000).
In comparison, the physical parameters of transient brightenings 
cover electron temperatures of $T_e \approx 4-8$ MK,
electron densities of $n_e \approx 2 \times 10^9$ cm$^{-3} - 
2 \times 10^{10}$ cm$^{-3}$, (loop) length scales of
$L = 5 \times 10^3$ km to $4 \times 10^4$ km, 
durations of $T = 2-7$ min,
and thermal energies of $10^{26}-10^{29}$ erg (Shimizu 1995).

Other studies yield similar values (Krucker \& Benz 1998;
Benz \& Krucker 2002; Parnell \& Jupp 2000), Li et al.~2012, in Fig.7 therein), 
but cannot be compared with the 3-D standard FD-SOC model here,
since they use a 2-D model that results into steeper 
power law slopes of $\alpha_E \gapprox 2.3-2.6$ for the 
flare energy. Generally, the thermal energy is defined by
\begin{equation}
	E_{th} = 3 k_B T_e n_e V = 3 k_B T_e
	\sqrt{EM\times V\times f} \ ,
\end{equation}
where $k_B$ is the Boltzmann constant, 
$EM = n_e^2 V$ is the volumetric emission measure, 
$V$ is the flare volume, and $f$ is a filling factor. 
In early studies the volume  has been quantified
by a flare area $A$ and a constant line-of-sight
depth $h_0$, i.e., $V=A\ h_0$, which corresponds to a
2-D (flat-world) model. Moreover, the fractality
of the spatial inhomogeneity is ignored with a
constant line-of-depth $h_0$. Hence, a definition
that is more consistent with the FD-SOC model is,
after inserting of $n_e = \sqrt{EM \times V}$,
\begin{equation}
	E_{th} = 3 k_B T_e n_e V = 3 k_B T_e
	\sqrt{EM\times L^{D_V}} \ ,
\end{equation}
where $D_V$ is the fractal (Hausdorff) dimension
and has a mean value of $D_V=2.5$ in a 3-D world.  

\subsection{	Hard X-ray Nonthermal Energies 	}

Size distributions of solar non-thermal flare 
energies have been sampled from ISEE-3/ICE
(Lu et al.~1993), and from HXRBS/SMM data 
(Crosby et al.~1993). The latter data set contains 
a number of $N_{\rm HXRBS}=11,352$ flare events and
has been accumulated over an observational epoch of 
${T_{\rm HXRBS} \approx 11}$ [years] (Crosby et 
al.~1993; Aschwanden et al.~2017).
The total energy in electrons ($>25$ keV) is found 
to cover a range of $E_{\rm HXR}\approx 3 \times
10^{28} - 10^{31}$ erg, as computed from 
the non-thermal electron energy radiated in hard X-ray 
(HXR) wavelengths at $\ge 25$ keV with the 
HXRBS detectors onboard the SMM spacecraft
(Dennis et al.~1985), assuming the thick-target 
bremsstrahlung model (Brown 1971). 
The energy in nonthermal electrons for the thick-target
model is given by (Brown 1971; Crosby et al.~1993),
\begin{equation}
	F(>E_0) = 4.8 \times 10^{24} A E_0^{-\gamma+1}
	E_m^\gamma \gamma (\gamma-1) 
	\beta(\gamma-{1\over 2},{1\over 2})
	\quad [{\rm erg}\ {\rm s^{-1}}] \ ,
\end{equation}
where $\beta$ is the beta function, and $E_0$ is the 
cut-off energy.
The power law slopes are found to be 
essentially invariant during the solar cycle, 
with best-fit values of 
$\alpha_E=1.53\pm0.02$, $1.51\pm0.04$, $1.48\pm0.02$, 
and $1.53\pm0.02$ during the year ranges of
1980-1982, 1983-1984, 1985-1987, and 1988-1989,
which can be compared with the theoretical FD-SOC 
model prediction of $\alpha_E=(5/3)\approx 1.67$.

\subsection{	White-Light Bolometric Energies	}

Homogeneous searches for stellar flares in every 
available KEPLER light curve (Borucki et al. 2010) 
revealed up to 851,168 candidate flare events, 
which have an average flare energy of $\approx 10^{35}$
erg (Davenport 2016). A KEPLER flare catalog with a 
sample of 162,262 events that supposedly controls
false-positive signals and artifacts of earlier
data sets revealed a power law slope of $\alpha_E
\approx 2$, from which a superflare with an energy
of $\approx 10^{34}$ was estimated to occur on the
Sun at least once in 5500 [years] (Yang \& Liu 2019).
For stellar flares we use the KEPLER data set of G-type
stars, which is most similar to the solar flare data set,
with the Sun being classified as a G5-type star. This 
stellar data set comprises $N_{\rm KEPLER}=55,269$ stellar 
flares and is observed during a time period of $T_{\rm KEPLER}
\approx 5$ [years] (Aschwanden \& G\"udel 2021). 
$E_{\rm bol}$ represents the energy of the 
bolometric (white-light) luminosity in the wavelength 
range of $\lambda \approx 4300-8900$ \ang\ with the 
KEPLER instrument. 
The wavelength range of KEPLER ($4300-8900$ \ang )
overlaps with the green line ($5548-5552$ \ang ),
the red line ($6682-6686$ \ang ), and the blue line 
($4502-4506$ \ang ). 

Following Namekata et al.~(2017a), there are
two differences between solar and stellar observations:
time cadences and passbands. HMI/SDO observes the
overall Sun with a 45 s cadene and a narrowband
filtergram around an Fe $\lambda$ 6173.3 line,
while KEPLER carried out 1-minute cadence observations
with 4000$-$9000 \ang\ broadband filters. The
white-light solar flare energy $E_{wl}$ is radiated by
a temperature $T_{flare}=10,000$ K, and is calculated as
\begin{equation}
	E_{wl} = \sigma_{\rm SB} T_{flare}^4 \int A_{flare}(t) dt \ ,
\end{equation}
with the stellar flare area $A_{flare}$ defined by,
\begin{equation}
	A_{flare}(t) = {L_{flare} \over L_{Sun}}
	\pi R^2
	{\int R_{\lambda} B_{\lambda}(5800\ {\rm K}) \ d\lambda
	\over
	R_{\lambda} B_{\lambda}(T_{flare})d\lambda } \ ,
\end{equation}
where $\sigma_{\rm SB}$ is the Stefan-Boltzmann constant,
$L_{flare}/L_{Sun}$ is the flare luminosity ratio to the
overall luminosity, $R$ is the solar radius,
$R_{\lambda}$ is a response function of HMI/SDO,
and $G_{\lambda}(T)$ is the Planck function at a
given wavelength $\lambda$. 
	
\section{	Theory of the  FD-SOC Model   }

In the following we describe briefly a definition of the 
{\sl Fractal-Diffusive Self-Organized Criticality (FD-SOC)} model,
while more extensive descriptions can be found in textbooks and
review papers (Aschwanden 2014, 2015, 2022, 2025; Aschwanden et al.~2016).
The FD-SOC model is based on four fundamental assumptions:
(i) the 3-D dimensionality, (ii) the mean fractal (Hausdorff) dimension,
(iii) the scale-freeness probability distribution function,
and (iv) the volume-flux proportionality for incoherent emission
processes.

Instead of using the next-neighbor interactions of cellular
automata, as defined in the original SOC model of Bak (1987),
we quantify the spatial inhomogeneity in terms of
the fractal dimension $D_d$ for the Euclidean domains
$d=1$ (lines), $d=2$ (areas), or $d=3$ (volumes).
Each fractal domain has a maximum fractal dimension of $D_d=d$,
a minimum value of $D_d=(d-1)$, and a mean value of $D_V=d-1/2$,
\begin{equation}
        D_V={(D_{\rm V,max} + D_{\rm V,min}) \over 2} = d-{1 \over 2}  \ .
\end{equation}
For most applications in the (observed) 3-D world, the dimensional
domain $d=3$ is appropriate, which implies a fractal dimension
$D_V=2.5$. However, if 2-D areas are observed, the fractal
dimension is $D_A=1.5$ and the dimensionality is $d=2$.
In this work we will mostly make use of the 3-D fractal domain,
while the 2-D domain is discussed elsewhere (e.g., Aschwanden 2022).
The fractal volume $V$ is then defined by the standard
(Hausdorff) fractal dimension $D_V$ in 3-D and the length scale
$L$ (Mandelbrot 1977),
\begin{equation}
        V(L) \propto L^{D_V} \ .
\end{equation}
We formulate the statistics of SOC
avalanches in terms of size distributions (or occurrence frequency
distributions) that obey the scale-free probability distribution
function (Aschwanden 2014, 2015, 2022), expressed with the power
law function
\begin{equation}
        N(L)\ dL \propto L^{-d} dL \ ,
\end{equation}
where $d=1,2,3$ represent the Euclidean dimensions
of the fractal domains
and $L$ is the length scale of a SOC avalanche. From this
scale-free relationship, the power-law slopes $\alpha_x$
of other SOC parameters $x=[A,V,F,E,T]$ can be derived,
such as for the area $A$,
the volume $V$, the flux $F$, the fluence or energy $E$, and
the duration $T$. The resulting power-law slopes $\alpha_x$
can then be obtained mathematically by the method of variable
substitution $x(L)$, by inserting the inverse function
$L(x)$ and its derivative $|dL/dx|$,
\begin{equation}
        N(x) dx = N[L(x)] \left| {dL \over dx} \right| dL
        = \ x^{-\alpha_x} dx \ ,
\end{equation}
such as for the flux $x=F$,
\begin{equation}
        \alpha_F = 1 + {(d-1) \over D_V \gamma} = {9 \over 5} = 1.80 \ ,
\end{equation}
or for the fluence $x=E$,
\begin{equation}
        \alpha_E = 1 + {(d-1) \over d \gamma} = {5 \over 3} = 1.67 \ .
\end{equation}
$\gamma$ is the nonlinearity coefficient in the flux-volume
relationship,
\begin{equation}
        F \propto V^\gamma = \left( L^{D_V} \right)^\gamma \ ,
\end{equation}
which degenerates to proportionality $F \propto V$ for $\gamma=1$.
Note that the definitions of the peak flux $F$ and fluence $E$ 
differ only by the fractal dimension $D_V$ (Eq.~10) and the
(space-filling) Euclidean dimension $d$ (Eq.~11). 
In astrophysical high-temperature plasmas, the volume $V$ is 
approximately proportional to the number of electrons in a region 
of instability, while the flux $F$ is proportional to the number 
of emitting photons, which implies that the photon-to-electron 
ratio is approximately constant and justifies the assumption of 
the flux-volume proportionality, in the case of incoherent 
emission mechanisms.

While this brief derivation of
Eqs.~(5-11) expresses the main assumptions of fractality
and linear flux-volume relationship for $\gamma=1$,
an additional assumption needs to be brought in that takes
the spatio-temporal evolution into account,
which can be approximated with the assumption 
of (classical) diffusive transport,
\begin{equation}
        L \propto T^{\beta/2} = T^{1/2} \ ,
\end{equation}
with the transport coefficient $\beta=1$. We call this
model the {\sl standard fractal-diffusive self-organized
criticality (FD-SOC)} model,
defined by [$d=3, \gamma=1, \beta=1$], while the
generalized FD-SOC model allows for variable coefficients
[$d, \gamma, \beta$] and alternative dimensionalities ($d = 1, 2$). 

\section{	DISCUSSION	}

Synthesized power law distributions have been constructed
earlier, such as from nanoflares to large solar flares
(Aschwanden et al.~2000), from thermal to nonthermal energies
(Li et al.~2012), for magnetic energies from bright points
to large active regions (Parnell et al.~2009),
from solar and stellar data (Shibayama et al.~2013),
from bolometric energies based on solar, stellar, lunar, and
terrestrial records (Schrijver et al.~2011, 2012). What is
new here is the application of the FD-SOC model into the
solar-stellar flare statistics, which predicts flare energies 
over an unprecedented scaling range of 13 orders of magnitude.

\subsection{	Synthesized Solar-Stellar Size Distribution }

We carry out a synthesized size distribution (Fig.~1) that can be
characterized with a power law function,
\begin{equation}
	N(E) dE = N_0[t, T] \left({E \over E_0}\right)^{-\alpha_E} dE \ ,
\end{equation} 
where $E$ is the total flare energy (in units of erg), 
$\alpha_E$ is the power law slope,
$E_0$ is an arbitrary reference value, or a threshold value, 
$N_0(t, T)$ is a normalization constant that
depends on the start time $t$ and duration $T$ 
of the observations in the data sampling.

The construction of such a synthesized size distribution is
shown in Fig.(1), stringed together from 5 different data
sets with different flare energy ranges. A total of 15
data sets are listed in Table 1, characterized with their energy
ranges $[E_{min}, E_{max}]$ and power law slopes $\alpha_E$,
from which only 5 data sets are represented in the
synthesized size distribution, while the other 10 cases
have some inconsistencies as described in the observation
Section 2.1.

What Fig.(1) shows is that each of the 5 displayed 
data sets exhibit a power law distribution that is 
consistent with the theoretical FD-SOC model with a slope 
of $\alpha_E=(5/3)\approx 1.67$. All 5 cases show also a
turnover at the lower end of the size distributions,
which is well-understood in terms of incomplete sampling
below the detection threshold. The term ``incomplete''
refers to the scale-free probability property, which
implies a reciprocality between the event frequency
$N(E)$ and the SOC parameter $E$ (Eq.~8). 

The graphical
representation in Fig.~(1) serves mostly to illustrate
the invariance of power law slopes $\alpha_E$ in each
data set, while the flaring frequency ($N(E)$, Eq.~14) 
is subject to different start times $t$, observational
duration $T$, and differences in the intrinsic energy 
ratios $q=E_1/E_2$. For instance, energy closure of flare
energies demonstrated ratios of 
$q=E_{elec}/E_{magn}=0.51\pm0.17$ for electron energies
between nonthermal electron energies and dissipated magnetic
energies, $q=E_{ions}/E_{magn}=0.17\pm0.17$ for ion energies,
$q=E_{CME}/E_{magn}=0.07\pm0.14$ for coronal mass ejection
(CME) energies, and
$q=E_{heat}/E_{magn}=0.07\pm0.17$ for direct heating processes,
which added together fulfill energy closure,
$q=E_{sum}/E_{magn}=0.87\pm0.18$ for the sum of all energies.
This suggests that a realistic evaluation of flare energies 
requires the summing of partial flare energies, such as 
thermal energies, non-thermal energies, and bolometric 
radiation energies (Aschwanden et al.~2017),
\begin{equation}
	E_{tot} = E_{nt} + E_{th} + E_{bol} \ .
\end{equation} 
For solar flares, the total radiative energy 
$E_{\rm bol}$ in visible
wavelengths amount to about 70\% of the total flare
energy, characterized by a blackbody temperature of
$T_{\rm bol} \approx 9000$ K (Woods et al.~2004).
Unfortunately, the complete energy budget and flare
partition is not known for most flare energetics studies.
Once all energy parameters are measured, a complete size
distribution $N(E_{tot})$ may be inferred that has the 
synthesized distribution as shown in Fig.~(1).

\subsection{    Flux and Fluence Limits  }

The flux is defined in physical units of [energy/time],
usually measured at the (background-subtracted) peak flux
$F$ at the peak time $t_p$ of an event,
\begin{equation}
        F={\rm max} \left[ f(t=t_p) \right] - f_{\rm BG} \  ,
\end{equation}
bound by the time range $t_1 \le t_p \le t_2$,
where $f(t)$ is the time profile of the flux.

The fluences have the physical unit of [energy],
and are measured by the time integral of the
time profile $f(t)$. Thus, the fluences $(E)$ are
the time-integrated fluxes,
\begin{equation}
        E = \int_{t_1}^{t_2} [f(t) - f_{\rm BG}] 
	\ dt \approx F \times T \ ,
\end{equation}
for a time interval ($t_1,t_2$) with well-defined start
times $t_1$ and end times $t_2$).
Combining the fluence approximation $E \approx F \times T$
(Eq.~17) with the flux-volume relationship $F \propto V$ (Eq.~11),
and the fractal dimension relatonship $V \propto L^{D_V}$ (Eq.~6),
we obtain the following scaling law for the length scale $L$
as a function of the flare energy $E$, with $D_V=5/2$,
\begin{equation}
	L(E) = E^{1/(2+D_V)} =  E^{2/9} \approx E^{0.22} \ .  
\end{equation}
which yields the following limit for an energy range of
13 orders of magnitude,
\begin{equation}
	\left( {L_{max} \over L_{min}} \right)  = 
	\left( {E_{max} \over E_{min}} \right)^{(2/9)}  = 
	\left( 10^{13} \right)^{(2/9)} \approx 774 \ .
\end{equation}
The predicted length scales for each order of magnitude
from 1 to 13 is enumerated in Table 2.
Using the scale of solar nanoflares as an estimation of
the smallest SOC event, i.e., $L_{min} \approx 1$ Mm
(Aschwanden et al.~2000; Cirtain et al.~2013), 
we would expect a maximum scale of 
$L_{max}=L_{min} \times 774$ Mm $\approx
1.04\ R_{\odot}$ for stellar flares. 
Leaving out the stellar flares we have an energy range of
9 orders of magnitude, which yields a range of 
$L_{max}=L_{min} \times 100$ Mm $\approx 0.14\ R_{\odot}$, 
which limits the maximum flare size to about the largest 
active region size, which represents a reasonable
scaling, given the observed sizes of the largest sunspots
and starspots. Schrijver (2011) estimates the size of the 
largest stellar flares to $L_{max} \approx (0.7-1.6) R_{\odot}$.
Since flare size distributions allow us to extrapolate 
and predict the probability of the most extreme events, 
the magnitude limit of these events depends also on the 
detection of an exponential drop-off of their size distribution.
For solar flares it appears that this finite-size effect
occurs at $L_{max} \lapprox 200$ Mm (Aschwanden \& Alexander 2001).

\subsection{	Constraints on Physical Processes	}

If a power law size distribution function extends over
multiple size distributions, many authors conclude that
a single physical process is responsible for the resulting
size distributions in multiple data sets. 
The finding of power law size distributions is then 
attributed to a single universal physical process. For instance,
finding a power law distribution of solar magnetic fields
over more than five decades in flux, Parnell et al.(2009) 
suggest that all surface magnetic features are generated
by the same physical mechanism, or that they are
dominated by surface processes (such as fragmentation,
coalescence, and cancellation) in a way which leads to
a scale-free distribution. 
Because of the similarity of their size distributions,
both solar and stellar flares are thought to be produced 
by a magnetic reconnection process (Pettersen 1989). 

From our identification of physical processes in our
samples we know that at least 3 different physical
processes are at play, namely (i) thermal free-emission, 
(ii) nonthermal thick-target bremsstrahlung, and (iii)
white-light bolometric emission. In addition,
conductive and radiative losses are thought to
contribute significantly to the dissipated flare
energy budget (Shimizu et al.~1995).
So, we have evidence
for at least 5 physical processes, although the
syntheseized size distribution function fits a sinlge
power law distribution over 13 order of magnitudes.
Thus, the invariant power law range is a necessary
but not a sufficient condition for an universal 
physical process.

How can we remedy this dilemma of energy partition.
Theoretically, we can measure the size distributions
of energies separately for each energy partition
and we can define the avalanche energy in terms of 
the total (summed) energy
dissipated during a flare. Care needs to be exercised
to avoid double counting of energies. For instance
an electron that gets accelerated in a flare plasma,
may also show up in the distribution of electrons
that produce free-free bremsstrahlung. Nevertheless,
energy closure in solar and stellar flares is feasible
(Aschwanden et al.~2017).

\section{	CONCLUSIONS  				}  

In this study we arrived at the following results and
conclusions:

\begin{enumerate}
\item{We construct a synthesized power law distribution
function over an energy range of $[E_{min}, E_{max}]
=[10^{24},10^{37}]$ [erg] that is consistent with
the theoretically predicted power law slope of 
$\alpha_E=(5/3) \approx 1.67$ of the {\sl fractal-diffusive 
self-organized criticality (FD-SOC)} model (Fig.~1).
The selected data sets (Table 1) encompass nanoflares,
microflares, transient brightenings, large solar flares
and stellar superflares. A subset of 5 selected
data sets (out of 15) is consistent with the 3-D 
FD-SOC model, which are used in the synthesized 
size distribution. Most of the remaining 5 data sets
use a 2-D model for the thermal flare energy
that we consider to be unrealistic, given the
numerous stereoscopic observations with the
two STEREO spacecraft. This is a case where
the {\sl Occham's razor} philosophy (parsimony law)
does not lead to a better model choice.}   

\item{The observed 5 size distributions all exhibit
a power law tail (AKAS inertial range) at the upper 
energy side, while the size distribution show 
a roll-over at the lower energy side, which is
readily understood as incomplete sampling in
a scale-free size distribution. This roll-over
is caused by a threshold value in the detection
of SOC avalanches.} 

\item{The observed length scales $L$ agree
with the theoretical prediction of the FD-SOC
model, i.e., $L \propto E^{2/9} \approx E^{0.22}$.
This scaling law has a weak energy dependence
that produces a small range of 2 orders of magnitude
for solar flares length scales $L$, 
i.e., 1 Mm $\le$ L $\le$ 100 Mm for
solar flares, and a range of 3 order of magnitudes
for stellar flares, i.e., 
1 Mm $\le$ L $\le$ 774 Mm = 1.04 $R_{\odot}$,
while the flare energy extends over 13 order of
magnitudes, i.e., 
$10^{24}$ [erg] $\le$ E $\le$ $10^{33}$ [erg].
Thus the FD-SOC model allows us a prediction 
of the SOC avalanche length scale $L$ for
any given flare energy $E$.}

\item{In the overall we conclude that the 
universality of power laws (and their slopes) 
is a consequence of SOC properties (fractality,
classical diffusion, scale-freeness, volume-flux
proportionality), rather than identical physical
processes at different wavelengths.} 

\end{enumerate}

Future studies may focus on the energy partition of
solar flare events, so that the total energy can be
estimated, since thermal, nonthermal, and white-light
bolometric energies appear to represent only lower 
limits of SOC avalanches only. 

\acknowledgments
{\sl Acknowledgements:}
This paper is dedicated to Carolus J.Schrijver (1958-2024)
in memoriam, who contributed substantially to this study.
We acknowledge constructive and stimulating discussions
(in alphabetical order)
with Arnold Benz, Sandra Chapman, Paul Charbonneau, 
Henrik Jeldtoft Jensen, Adam Kowalski, Sam Krucker, 
Alexander Milovanov, Leonty Miroshnichenko, Jens Juul 
Rasmussen, Carolus Schrijver, Vadim Uritsky, Loukas Vlahos, 
and Nick Watkins.
This work was partially supported by NASA contract NNX11A099G
``Self-organized criticality in solar physics'' and NASA contract
NNG04EA00C of the SDO/AIA instrument to LMSAL.

%\clearpage

\section*{      References      }
\def\ref#1{\par\noindent\hangindent1cm {#1}}

\ref{Aschwanden, M.J., Tarbell, T.D., Nightingale, W.,
	Schrijver, C.J., Title A., Kankelborg, C.C., and Martens, P. 2000,
	{\sl Time variability of the ``quiet'' Sun observed with TRACE.
	II. Physical parameters temperature evolution, and energetics
	of extreme-ultraviolet nanoflares}, ApJ 535, 1047-1065.}  
\ref{Aschwanden, M.J. and Alexander, D. 2001,
	{\sl Flare plasma cooling from 30 MK down to 1 MK modeled
	from Yohkoh, GOES, and TRACE observatons during the
	Bastille day event (14 July 2000)}, SoPh 204: 93-121.}
\ref{Aschwanden, M.J. and Parnell, C.E., 2002,
	{\sl Nanoflare statistics from first principles:
	Fractal geometry and temperature synthesis},
	ApJ 572, 1048-1071.}

\ref{Aschwanden, M.J. 2004, {\sl Physics of the solar corona.
	An Introduction}, Springer PRAXIS, Berlin, Heidelberg.}
\ref{Aschwanden, M.J. 2014,
        {\sl A macroscopic description of self-organized systems and
        astrophysical applications}, ApJ 782, 54.}
\ref{Aschwanden, M.J. 2011,
        {\sl Self-Organized Criticality in Astrophysics. The Statistics
        of Nonlinear Processes in the Universe},
        Springer-Praxis: New York, 416p.}
\ref{Aschwanden, M.J. 2015,
        {\sl Thresholded power law size distributions of instabilities
        in astrophysics}, ApJ 814.}
\ref{Aschwanden, M.J. 2013a, in {\sl Theoretical Models of SOC Systems},
        chapter 2 in {\sl Self-Organized Criticality Systems}
        (ed. Aschwanden M.J.), Open Academic Press: Berlin, Warsaw, p.21.}
\ref{Aschwanden,M.J. 2013b,
        {\sl Self-Organized Criticality Systems in Astrophysics (Chapter 13)},
        in "Self-Organized Criticality Systems" (ed. Aschwanden,M.J.),
        Open Academic Press: Berlin, Warsaw, p.439.}
\ref{Aschwanden, M.J., Crosby, N., Dimitropoulou, M., Georgoulis, M.K.,
        Hergarten, S., McAteer, J., Milovanov, A., Mineshige, S.,
        Morales, L., Nishizuka, N., Pruessner, G., Sanchez, R.,
        Sharma, S., Strugarek, A., and Uritsky, V. 2016,
        {\sl 25 Years of Self-Organized Criticality: Solar and
        Astrophysics}, Space Science Reviews 198, 47.}
\ref{Aschwanden, M.J., Caspi, A., Cohen, C.M.S., Gordon, H.,
	Jing, J., Kretzschmar, M. et al.~2017,
	{\sl Coronal energetics of solar flares. V. Energy closure
	and coronal mass ejections}, ApJ 836/1, article id.17, 17pp.}
\ref{Aschwanden, M.J. 2021,
        {\sl Finite system-size effects in self-organizing criticality 
	systems}, ApJ 909, 69.}
\ref{Aschwanden, M.J. and Guedel, M. 2021, {\sl Self-organized
	criticality in sellar flares}, ApJ 910, id.41, 16pp.}
\ref{Aschwanden, M.J. 2022,
        {\sl The fractality and size distributions of astrophysical
        self-organized criticality systems},
        ApJ 934 33}
\ref{Aschwanden, M.J. and Gogus, E. 2025,
	{\sl Testing the universality of self-organized criticality
	in galactic, extra-galactic, and black-hole systems},
	ApJ, 978:19 (11pp).}
\ref{Aschwanden, M.J. 2025,
	{\sl Power Laws in Astrophysics. Self-Organzed Criticality
	Systems}, Cambridge University Press: Cambridge.}
\ref{Bak, P., Tang, C., and Wiesenfeld, K. 1987,
        {\sl Self-organized criticality: An explanation of 1/f noise},
        Physical Review Lett. 59(27), 381.}
\ref{Bak, P. 1996,
        {\sl How Nature Works. The Science of Self-Organized Criticality},
        Copernicus: New York.}
\ref{Benz, A.O. and Krucker, S. 2002,
        {\sl Energy distribution of microevents in the quiet solar corona},       
        ApJ 568, 413-421.}
\ref{Borucki, W.J., Koch, D., Gibor, B., Batalha, N.B., Brown, T., 
	et al.~2010, {\sl Kepler planet-detection mission: Introduction
	and first results}, Science 327, Issue 5968, 977.}
\ref{Bromund, K.R., McTiernan, J.M., Kane, S.R. 1995,
        {\sl Statistical studies of ISEE3/ ICE observations of impulsive
        hard X-ray solar flares}, 
        ApJ 455, 733-745.}
\ref{Brown, J.C. 1971, {\sl The deduction of energy spectra of non-thermal
	electrons in flares from the observed dynamic spectra of hard X-ray
	bursts}, SoPh 18, 489.}
\ref{Cai, Y., Hou, Y., Li, T., Liu, J. 2004,
	{\sl Statistics of solar white-flight flares. I. Optimization and
	application of identification methods}, ApJ 975:69 (16pp).} 
\ref{Christe, S., Hannah, I.G., Krucker, S., McTiernan, J., and Lin, R.P. 2008, 
        {\sl RHESSI microflare statistics. I. Flare-finding and
        frequency distributions}, 
        ApJ 677, 1385-1394.}
\ref{Crosby, N.B., Aschwanden, M.J., and Dennis, B.R. 1993,
	{\sl Frequency distributions and correlations of solar flare
	parameters}, SoPh 143, 275-299.}
\ref{Davenport, J.R.A. 2016, {\sl The Kepler catalog of stellar flares},
	ApJ 829:23 (12pp).}
\ref{Dennis, B.R. 1985,
        {\sl Solar hard X-ray bursts}, 
        Sol.Phys. 100, 465-490.}
\ref{Hannah, I.G., Christe, S., Krucker, S., Hurford, G.J., Hudson, H.S.
	Lin P.R. 2008, {\sl RHESSI microflare statistics. II. X-ray
	imaging, spectroscopy, and energy distributions},
	ApJ 677:704-718.}
\ref{Krucker, S. and Benz, A.O. 1998,
        {\sl Energy distribution of heating processes in the quiet solar
        corona}, ApJ 501, L213.}
\ref{Li, Y.P., Gan, W.Q., and Feng, L. 2012,
	{\sl Statistical analyses on thermal aspects of solar flares},
	ApJ 747, 133 (8pp).}
\ref{Lu, E.T., Hamilton, R.J., McTiernan, J.M., and Bromund, K.R. 1993,
        {\sl Solar flares and avalanches in driven dissipative systems},
        ApJ 412, 841-852.}
\ref{Mandelbrot, B.B. 1977,
        {\sl Fractals: form, chance, and dimension}, Translation of
        {\sl Les objects fractals}, W.H. Freeman, San Francisco.}
\ref{McAteer,R.T.J., Aschwanden,M.J., Dimitropoulou,M., Georgoulis,M.K.,
        Pruessner, G., Morales, L., Ireland, J., and Abramenko,V. 2016,
        {\sl 25 Years of Self-Organized Criticality: Numerical Detection
        Methods}, SSRv 198, 217.}
\ref{Namekata, K., Sakaue, T., Watanabe, K., Asai, A., Shibata, K.,
	et al. 2017a, {\sl Statistical studies of solar white-light
	flares and comparisons with superflares on solar-type stars},
	ApJ 851/2, id.91, 14pp.}
\ref{Namekata, K., Sakaue, T., Watanabe, K., Asai, A., and Shibata, K. 2017b,
	{\sl Validation of a scaling law for the coronal magnetic field
	strength and loop length of solar and stellar flares},
	PASJ 69/1, id.7 10pp.}
\ref{Parnell,C.E. and Jupp,P.E. 2000,
        {\sl Statistical analysis of the energy distribution of nanoflares
        in the quiet Sun}
        ApJ 529, 554-569.}
\ref{Parnell, C.E., DeForest, C.E., Hagenaar, H.J., Johnston, B.A., Lamb, D.A.,
        and Welsch, B.T. 2009,
        {\sl A Power-Law Distribution of Solar Magnetic Fields Over More Than     
        Five Decades in Flux},
        ApJ 698, 75-82.}
\ref{Paudel, R.R., Gizis, J.E., Mullan, D.J., Schmidt, S.J.,
	Burgasser, A.J. and Williams, P.K.G. 2020, {\sl K2 Ultra dwarfs survey -
	IV. Whie light superflares observed on an L5 dwarf and flare rates of 
	L dwarfs}, MNRAS 494, 5751-5760.}
\ref{Pettersen, B.R. 1989, {\sl A review of stellar flares and their
	characteristics}, SoPh 121/1-2, pp.299-312.}
\ref{Pruessner, G. 2012, {\sl Self-Organised Criticality. Theory, Models
        and Characterisation}, Cambridge University Press: Cambridge.}
\ref{Sharma,A.S., Aschwanden,M.J., Crosby,N.B., Klimas,A.J., Milovanov,A.V.,
        Morales,L., Sanchez,R., and Uritsky,V. 2016,
        {\sl 25 Years of Self-Organized Criticality: Space and Laboratory
        Plamsas}, SSRv 198, 167.}
\ref{Schrijver, C.J. 2011, {\sl Solar energetic events, the solar-stellar
	connection, and extreme space weather},
	(ed., Johns-Krull, C.M., Browning, M.K., and West,A.A.; 
	San Francisco: Astronomical Society of the Pacific), 
	ASP Conf.Ser. 448, p.231.}
\ref{Schrijver, C.J., Beer, J., Baltensperger, U., CLiver, E.W., et al. 2012,     
        {\sl Estimating the frequency of extremely energetic solar events,     	  
	based on solar, stellar, lunar, and terrestrial records},
        JGR (Space Physics) 117, 8103.}
\ref{Shibayama, T., Maehara, H., Notsu, Y., Nagao, T., Satoshi,, H.
	et al.~(2013), {\sl Superflares on solar-type stars observed with
	Kepler. I. Statistical properties of superflares},
	ApJSS 209/1, article id.5, 13pp.}
\ref{Shimizu, T. 1995, {\sl Energetics and occurrence rate of active region
	transient brightenings and implications for the heating of the
	active-region corona}, PASJ 47, 251-263.}
\ref{Watkins, N.W., Pruessner, G., Chapman, S.C., Crosby, N.B., and Jensen, H.J.
        {\sl 25 Years of Self-organized Criticality: Concepts and 
	Controversies}, 2016, SSRv 198, 3.}
\ref{Woods, T.N., Eparvier, F.G., Fontenla, J., Harder, G.,
	Kopp, W.E. et al. 2004, {\sl Solar irradiance variability
	during the October 2003 solar storm period},
	Geophys.Res.Let. 31, L10802.}
\ref{Yang, Y. and Liu, J. 2019, {\sl The flare catalog and the flare
	activity in the Kepler mission}, ApJSS 241:29 (19pp).}

\clearpage
%%%%%%%%%%%%%%%%%%%%%%%%%%%% TABLE 1 &&&&&&&&&&&&&&&&&&&&&
\begin{table}
\begin{center}
\caption{
The energy ranges [$E_{min}, E_{max}$] and power law slopes 
$\alpha_E$ of solar and stellar flare energies.}
\normalsize
\medskip
\begin{tabular}{llllll}
\hline
Object, Instrument                 & Energy range            & Power law     & References\\
                                   & (erg)                   & slope         &           \\
                                   & [$E_{min}, E_{max}$]    & $\alpha_E$    &           \\
\hline
\hline
{\bf Thermal energy $E_{Th}$:}	   &                         &               &               \\
Solar flares, SXT/Yohkoh 	   & [$10^{25}-10^{29}$]     & 1.5$-$1.6     & Shimizu (1995), Fig.11\\
Solar flares, EIT/SOHO		   & [$10^{24.9}-10^{26.3}$] & 2.3$-$2.6 
 & Krucker \& Benz (1998), Fig.2\\
Solar EUV nanoflares TRACE	   & [$10^{24}-10^{26}$]     & 1.79          & Aschwanden et al.(2000), Fig.10\\
Solar flares, TRACE		   & [$10^{23}-10^{26}$]     & 2.4$-$2.6     & Parnell \& Jupp (2000), Figs.1,4,5\\
Solar EUV nanoflares, EIT/SOHO	   & [$10^{25}-10^{27}$]     & 2.31          & Benz \& Krucker (2002), Figs.4,5,8\\
Solar flares, GOES		   & [$10^{24}-10^{32}$]     & 2.38          & Li et al. (2012), Fig.7\\
\hline
{\bf Non-thermal energy $E_{NT}$:} &                         &               &           \\
Solar flares, HXRBS/SMM            & [$10^{25}-10^{32}$]     & 1.53$\pm$0.02 & Crosby et al.(1993), Fig.6\\
Solar flares, ISEE-3/ICE           & [$10^{27}-10^{33}$]     & 1.47$\pm$0.04 & Lu et al.(1993), Fig.8\\ 
Solar flares, ISEE-3/ICE           & [$10^{29}-10^{30}$]     & 1.67$-$1.74   & Bromund et al.(1995), Table 1\\
Solar microflares, RHESSI	   & [$10^{24}-10^{30}$]     & 2.0           & Hannah et al.(2008), Fig. 18)\\
\hline
{\bf White-light radiative energy} $E_{wl}$: &               &               &                 \\
Stellar flares KEPLER KIC          & [$10^{33}-10^{36}$]     & 1.56, 1.98    & Davenport (2016), Fig.6\\
Stellar flares KEPLER G-type stars & [$10^{32}-10^{37}$]     & 1.96$\pm$0.04 & Yang \& Liu (2019), Fig.3\\
Stellar flares GOES class X,M,C    & [$10^{28}-10^{32}$]     & ...           & Cai et al.(2024), Fig.9\\
Stellar flares L dwarfs            & [$10^{33}-10^{34}$]     & ...           & Paudel et al.(2024), Fig.4 3\\ 
Stellar flares solar-type          & [$10^{29}-10^{36}$]     & ...           & Namekate et al.(2017a), Figs.6,8,12\\
\hline
{\bf Synthesized energy distributions:} &                    &               & \\
EUV + SXR + HXR                    & [$10^{24}-10^{32}$]     & 1.80          & Aschwanden et al.(2000), Fig.10 \\
EUV + SXR                          & [$10^{24}-10^{30}$]     & 1.54$\pm$0.03 & Aschwanden \& Parnell (2002), Fig.10 \\
TB + QS + AR                       & [$10^{24}-10^{30}$]     & 1.54$\pm$0.05 & Aschwanden (2004), Fig. 9.27   \\
QS + AR                            & [$10^{17}-10^{23}$ Mx]  & 1.85$\pm$0.14 & Parnell et al.~(2009), Fig.5   \\
EUV + SXR + HXR + SF               & [$10^{24}-10^{35}$]     & 1.87$\pm$0.10 & Schrijver et al.~(2011), Fig.3 \\
EUV + SXR + HXR + SF               & [$10^{24}-10^{37}$]     & 2.30          & Schrijver et al.~(2012), Fig.3 \\
EUV + SXR + HXR                    & [$10^{24}-10^{32}$]     & 2.0$-$2.3     & Li et al.~(2012), Fig.7        \\
EUV + SXR + HXR + SF               & [$10^{24}-10^{36}$]     & 1.80          & Shibayama et al.~(2013), Fig. 13.12 \\
\hline
\end{tabular}
\end{center}
\par EUV=extreme ultraviiolet, SXR=soft X-rays, HXR=hard X-rays,
TB=Transient brightenings, QS=Quiet Sun, AR=Active regions,
SF=Stellar flares.
\end{table}

%%%%%%%%%%%%%%%%%%%%%%%%%%%% TABLE 2 &&&&&&&&&&&&&&&&&&&&&
\begin{table}
\begin{center}
\caption{Predicted scaling law of length scales $L(E)$ 
as a function of the solar and stellar flare energies $E$.}
\normalsize
\medskip
\begin{tabular}{rrrrl}
\hline
Relative     & Absolute  & Relative      & Predicted     & Nomenclature \\
energy       & energy    & energy        & length scale  & \\
logarithm    & logarithm & factor        & $L(E)$ [Mm]   & \\
\hline
\hline
       0     &   24 &		       1 & 1.00 Mm & Nanoflares \\
       1     & 	 25 &     	      10 & 1.67 Mm \\
       2     &	 26 &                100 & 2.78 Mm \\
       3     &   27 &              1,000 & 4.64 Mm & Microflares \\
       4     &   28 &             10,000 & 7.74 Mm \\
       5     &   29 &            100,000 & 12.9 Mm \\
       6     &   30 &          1,000,000 & 21.5 Mm & Large solar flares \\
       7     &   31 &         10,000,000 & 35.9 Mm \\
       8     &   32 &        100,000,000 & 59.9 Mm & \\
       9     &   33 &      1,000,000,000 & 100  Mm & Stellar flares \\
      10     &   34 &     10,000,000,000 & 167  Mm & \\
      11     &   35 &    100,000,000,000 & 278  Mm & \\
      12     &   36 &  1,000,000,000,000 & 464  Mm & \\
      13     &   37 & 10,000,000,000,000 & 774  Mm & Stellar superflares \\
\hline
\end{tabular}
\end{center}
\end{table}

%%%%%%%%%%%%%%%%%%%%%%%%%%%%%% FIGURES %%%%%%%%%%%%%%%%%%%%%%%%%
\begin{figure}[t]
\centerline{\includegraphics[width=1.0\textwidth]{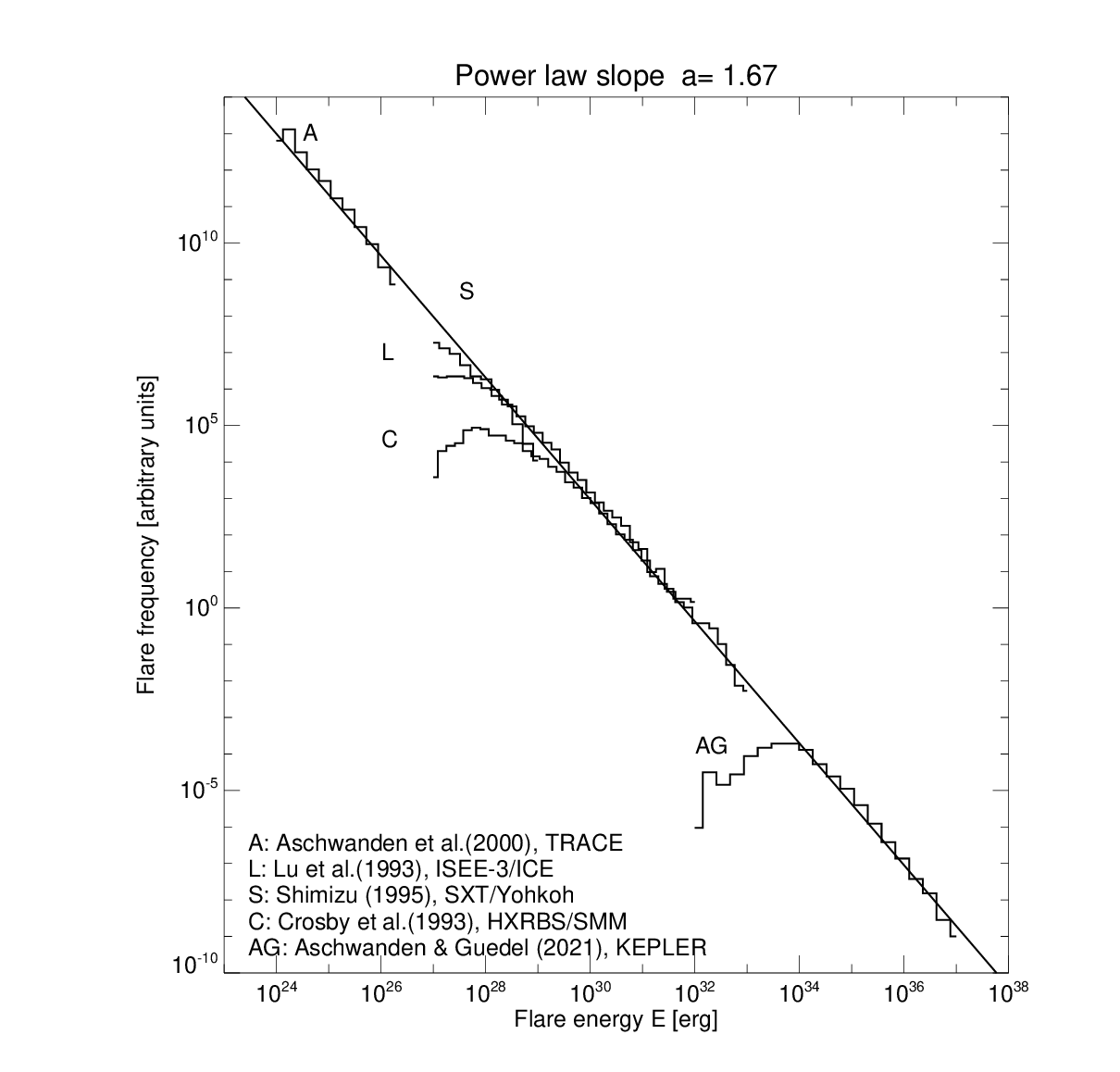}}
\caption{Energy size distributions synthesized from solar
observations with HXRBS/SMM, ISEE-3/ICE, SXT/Yohkon, TRACE,
and from stellar observations with KEPLER. The theoretical 
prediction of the power law slope for the flare energy 
distribution is $\alpha_E=1.67$ in the FD-SOC model.
The flare frequencies are aligned to the predicted power law
size distribution.}
\end{figure}

\end{document}